\documentclass[lettersize,journal]{IEEEtran}
\usepackage{amsmath,amsfonts}
\usepackage{algorithmic}
\usepackage{algorithm}
\usepackage{array}
\usepackage[caption=false,font=normalsize,labelfont=sf,textfont=sf]{subfig}
\usepackage{textcomp}
\usepackage{stfloats}
\usepackage{url}
\usepackage{verbatim}
\usepackage{graphicx}
\usepackage{cite}
\usepackage{hyperref}
\usepackage{xcolor}
\begin{document}

\title{Adversarial Learning-Based Radio Map Reconstruction for Fingerprinting Localization\thanks{This work was supported by the Leverhulme Trust under Research Leadership Award RL-2019-019(\textit{Corresponding author: Jiajun He})}
}

\author{Jiaming~Zhang,~\IEEEmembership{Student Member,~IEEE,}
        Jiajun~He,~\IEEEmembership{Member,~IEEE,}
        Tianyu~Lu,~\IEEEmembership{Member,~IEEE,} Jie~Zhang,~\IEEEmembership{Member,~IEEE,}
        Okan~Yurduseven~\IEEEmembership{Senior Member,~IEEE}}
\markboth{IEEE Antennas and Wireless Propagation Letters}%
{Shell \MakeLowercase{\textit{et al.}}: A Sample Article Using IEEEtran.cls for IEEE Journals}

\maketitle

\begin{abstract}
This letter presents a feature-guided adversarial framework, namely ComGAN, which is designed to reconstruct an incomplete fingerprint database by inferring missing received signal strength (RSS) values at unmeasured reference points (RPs). An auxiliary subnetwork is integrated into a conditional generative adversarial network (cGAN) to enable spatial feature learning. An optimization method is then developed to refine the RSS predictions by aggregating multiple prediction sets, achieving an improved localization performance. Experimental results demonstrate that the proposed scheme achieves a root mean squared error (RMSE) comparable to the ground-truth measurements while outperforming state-of-the-art reconstruction methods. When the reconstructed fingerprint is combined with measured data for training, the fingerprinting localization achieves accuracy comparable to models trained on fully measured datasets.  
\end{abstract}

\begin{IEEEkeywords}
Fingerprinting, generative adversarial network, internet-of-things, positioning, received signal strength.
\end{IEEEkeywords}

\section{Introduction}

\IEEEPARstart{P}{ositioning} using received signal strength (RSS) has been widely studied for applications such as indoor navigation, object tracking, surveillance, etc \cite{6786980, 9993721, 7047740}. Traditional multilateration methods~\cite{hcso-RSS} often suffer from significant errors due to multipath fading and environmental blockages. In contrast, fingerprinting algorithms \cite{11165772, 9542942} typically achieve a higher localization accuracy by formulating the localization problem as an RSS pattern-matching task. In this approach, RSS samples collected at reference points (RPs) during the offline phase are compared with online RSS observations to determine the target position. Nevertheless, collecting RSSs at all RPs is labor-intensive and time-consuming, particularly in large-scale environments. To address this challenge, recent studies have investigated expanding fingerprint databases using synthetic RSS.


Generative models have been studied for synthesizing RSS samples \cite{gao2025genmetaloc, 10025551, 10747207}. Among these approaches, generative adversarial networks (GANs) stand out for their efficiency, ability to generate RSS samples in a single pass, and well-established training ecosystem, making them practical for real-time and resource-constrained Internet of Things (IoT) applications. Nijma $et$ $al.$ \cite{9812625} applied a standard GAN framework to generate synthetic RSS values along with the corresponding RPs by feeding random vectors. Furthermore, several studies have explored the controllability of synthetic RSS sample generation at measured RPs. In \cite{boulis2021data}, a conditional GAN (cGAN) was presented, which used measured RPs as additional inputs with random vectors to generate spatially conditioned RSS samples, thereby reducing localization error. Junoh \textit{et al.} \cite{10443392} incorporated a long short-term memory (LSTM) module into the cGAN framework, aiming to capture time-dependent characteristics of RSS fluctuations. To enhance controllability in synthesizing RSS at unmeasured RPs, Zou $et$ $al.$ \cite{zou2020robot} proposed GPR-GAN, a cascaded architecture that integrates Gaussian process regression (GPR) with a GAN. In this framework, the GPR estimated the average RSS from RP inputs, while the GAN refined the estimated RSS measurements to capture the impacts of fine-grained fading and stochastic perturbations. Nevertheless, the performance of GPR-GAN is constrained by the limited expressiveness of GPR's predefined kernels, which struggled to model the complex spatial correlations presented in real RSS datasets \cite{7827145}.



To reconstruct RSS fingerprints at unmeasured RPs, we propose a generative framework, ComGAN, that leverages partial RSS measurements and RP information for data completion. The contributions of this work are threefold: 1) \textit{Learning-based RSS Prediction}: Developed a feature-guided adversarial framework that integrates an auxiliary RP estimation branch to enable spatial feature learning; 2) \textit{Loss Optimization for RSS Recovery}: Designed a hybrid loss function that jointly optimizes adversarial and RP-aware objectives for accurate and stable RSS reconstruction; and 3) \textit{Prediction Optimization Strategy}: Introduced a subset-aggregation strategy that refines predicted RSS values via random subset averaging, effectively reducing localization error while lowering fingerprinting data-collection costs.

\textit{Notation}:
Bold lower-case letters represent column vectors, while capital letters are matrices. The superscripts $(\cdot)^T$ and $(\cdot)^{-1}$ represent the transpose and inverse operations, respectively. The estimate of the variable $x$ is denoted by $\hat{x}$, $\|\cdot\|$ is the Euclidean norm, while $\odot$ denotes element-wise multiplication.

\section{Problem Formulation}
\label{sec:Problem Formulation}

This section begins by outlining the RSS model, followed by a brief description of the fingerprinting algorithm.


\subsection{RSS Measurement}
Considering a two-dimensional (2D) localization problem, our goal is to determine the target position $\mathbf{p} = [x \ y]^T$ with $L$ radio units (RUs). Let us denote the transmit power of the target (a mobile device) as $P_{t}$, the received signal power $P_{r,l}$ between the target and $l$-th RU ($\mathbf{p}_l$) is given by \cite{hcso-RSS}: 
\begin{equation}
    P_{r,l} = P_t K_l h_{l}d_{l}^{-\alpha},
\label{eq:power}
\end{equation}
where $d_l = \| \mathbf{p} - \mathbf{p}_l \|$; $K_l$ is the antenna gain; $h_{l}$ denotes the channel fading gain, while $\alpha$ is the path-loss exponent. By applying a natural logarithm to both sides of \eqref{eq:power}, the RSS measured at the $l$-th RU is then \cite{hcso-RSS}:
\begin{equation}
    r_{\text{RSS},l} = \ln P_t - \ln h_l - \ln K_l = -\alpha\ln(d_l) + n,
\end{equation}
where $n$ is the additive noise. However, in real-world scenarios, RSS measurements are often unstable due to the variability of signal propagation environments. The presence of Non-Line-of-Sight (NLoS) conditions and blockage effects can significantly distort RSS values, leading to considerable degradation in the performance of multilateration-based localization methods \cite{11165772, he-LTE}. To address this challenge, fingerprinting algorithms have emerged as a promising solution for target localization in mixed Line-of-Sight (LoS)/NLoS environments.

\vspace{-0.5cm}
\subsection{Fingerprinting Algorithm}

\begin{figure}[t]
    \centering
    \includegraphics[width=0.88\linewidth, height=0.3\linewidth]{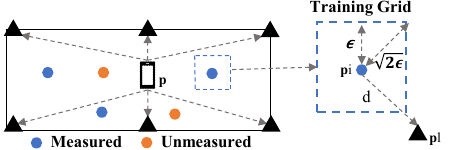}
    \caption{Illustration of the fingerprint localization problem.} 
    \vspace{-15pt}
    \label{fig:sysmodel}
\end{figure}

The fingerprinting algorithm consists of two major phases: 1) \textit{Offline Phase}: Collecting RSS samples at each RP to form the fingerprint database; and 2) \textit{Online Phase}: The newly acquired RSS measurements are compared with the database for target localization. Concretely, during the offline phase, a 2D localization area is partitioned into multiple training grids with a spacing of $\epsilon$, as illustrated in Fig. \ref{fig:sysmodel}. The centre of each grid ($\mathbf{p}_i = [x_i \ y_i]^T$ and $i=1,2,\ldots$) is treated as a RP for fingerprint collection. Given a centre point $\mathbf{p}_i$, the RSS vector observed at the $l$-th RU is given by:
\begin{equation}
    \mathbf{r}_{\text{RSS},i,l} = \left[r^{1}_{\text{RSS},i,l}, r^{2}_{\text{RSS},i,l}, \cdots, r^{N}_{\text{RSS},i,l}\right]^T,
\end{equation}
where $N$ represents the total number of RSS observations. By collecting RSS samples at all the RPs, a fingerprint database is constructed. Regarding the online phase, pattern matching is employed to estimate the target location using the observed RSS values, where the RP whose stored fingerprint exhibits the highest similarity to the newly observed RSS is identified as the target position. 

\section{GAN-based Fingerprint Completion}

In this section, the network architecture design is first presented, followed by the design of the loss function. Lastly, the proposed optimization strategy is introduced.

\label{sec:method}

\vspace{-0.5cm}
\subsection{Network Architectures}

Inspired by image inpainting \cite{guillemot2013image}, RSS completion aims to infer missing values, but requires a feature-guided framework to capture RP-dependent structures while preserving distributional fidelity. Thus, an adversarial scheme enforces distributional fidelity beyond element-wise reconstruction. This framework comprises two adversarially trained subnetworks: 1) a generator reconstructs complete RSS from incomplete inputs; and 2) a critic distinguishes real from predicted samples.

\begin{figure}[t]
    \centering
    \includegraphics[width=0.9\linewidth, height = 0.4\linewidth]{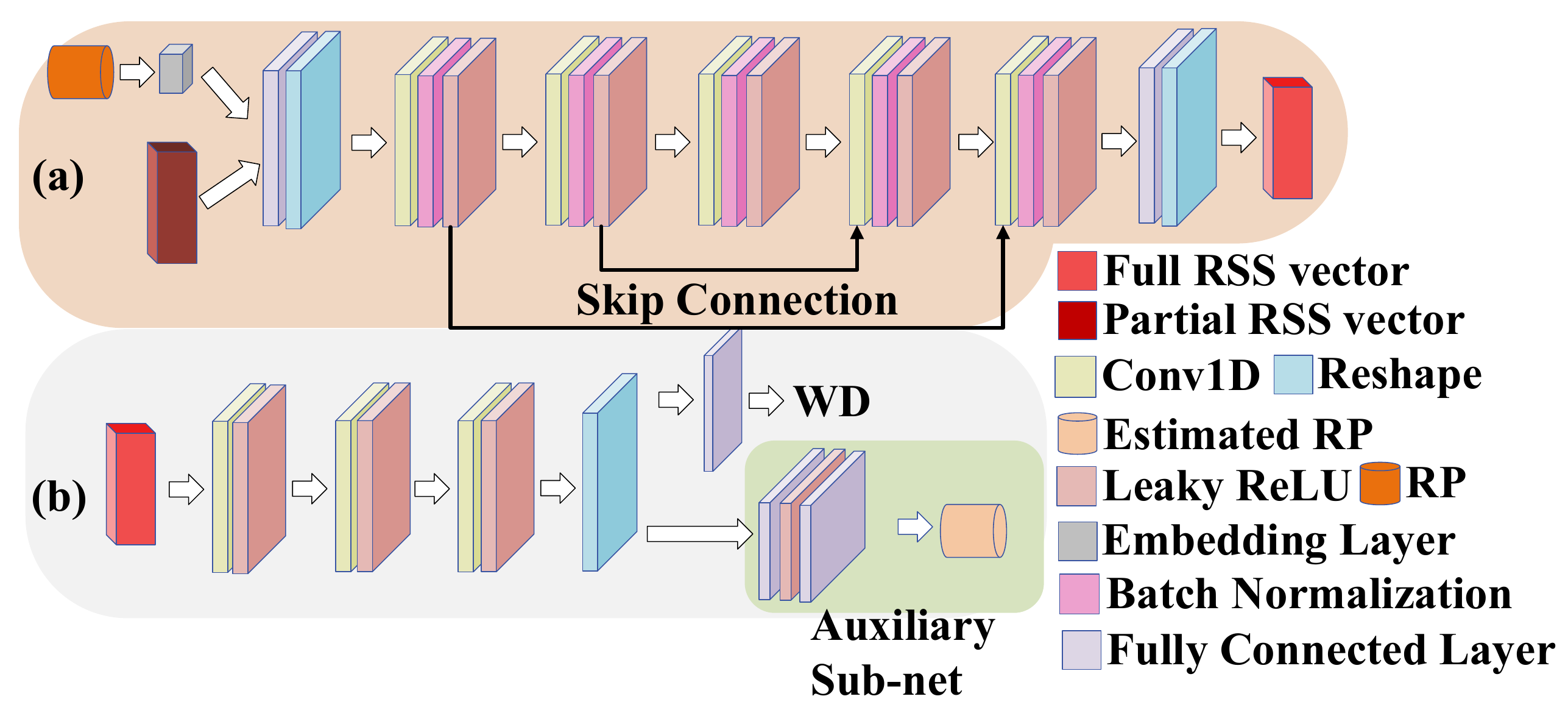}
    \caption{The architecture of ComGAN (a) generator and (b) critic, and the spatial distribution of RPs.} \vspace{-15pt}
    \label{fig:arc}
\end{figure}

As previously discussed, RSS values vary with the geo-location of the RP \cite{he2018SLAC, he2024BLM}. To model this dependency, both incomplete RSS samples and their corresponding RPs are used as inputs. As shown in Fig.~\ref{fig:arc}(a), the generator employs an embedding layer to align RP and RSS dimensions \cite{hrinchuk2019tensorized}. The embedded RP features are concatenated with RSS inputs and passed through a fully connected layer. Five computational blocks then process the resulting representations, each consisting of a one-dimensional convolutional (Conv1D) layer, batch normalization, and a leaky rectified linear unit (ReLU) activation \cite{9163446}. The Conv1D layers capture local correlations among RSS measurements, while skip connections preserve fine-grained details across layers. A final fully connected layer maps the features to the target dimensionality, generating the reconstructed fingerprint database.

The critic shares a similar backbone with the generator (Fig.~\ref{fig:arc}(b)) but extends its functionality. It takes either real or predicted RSS samples as input and outputs (i) a scalar authenticity score and (ii) an auxiliary RP estimation. The auxiliary branch, composed of two fully connected layers with a leaky ReLU activation, serves as a feature-guided regularization component that enforces stable feature learning. In summary, the generator performs RSS completion by leveraging RP information and incomplete RSS samples, whereas the critic improves prediction fidelity by discriminating predicted RSS and incorporating RP-aware, feature-guided semantic supervision via the auxiliary estimation branch.

\vspace{-0.5cm}

\subsection{Loss Function}
ComGAN adopts the Wasserstein score (WS), which provides smoother gradients and better stability, particularly beneficial for skewed, heavy-tailed RSS distributions caused by multipath fading and signal blockage \cite{9542942}. To mitigate the risk of predicted RSS samples being biased toward environmental noise, an RP-aware regularization objective, defined as the mean squared error (MSE) between two types of RP estimations, is incorporated into the critic loss to guide the model toward more stable and spatially consistent feature learning. Let $\hat{c}_1$ and $\hat{c}_2$ denote the estimated RPs when the input of the critic is real and predicted RSS samples, respectively. The overall loss function for the critic, denoted as $\mathcal{L}_{C}$, is given by:
\begin{equation}
\mathcal{L}_{C} = \mathcal{L}_{\text{C, adv}} + \lambda_1 \mathcal{L}_{\text{GP}} + \lambda_2 \text{MSE}(\hat{c}_1,\hat{c}_2),
\end{equation}
where $\mathcal{L}_\text{C, adv}$ is the WS, and $\mathcal{L}_{\text{GP}}$ denotes the gradient penalty (GP) for improved training stability \cite{gulrajani2017improved}; $\lambda_1$ and $\lambda_2$ are the weighting coefficients for the GP term and the RP estimation loss term, respectively. Following the recommendation in \cite{gulrajani2017improved}, $\lambda_1$ is set to 10. The coefficient $\lambda_2$ is empirically set to $1$. Since $\mathcal{L}_{\text{GP}}$ relates to the training of the critic rather than the generator, it is excluded from the generator’s loss function. The total loss function for the generator $\mathcal{L}_{G}$ is:
\begin{equation}
    \mathcal{L}_{G} = \mathcal{L}_{\text{G, adv}} + \lambda_2 \text{MSE}(\hat{c}_1,\hat{c}_2) + \lambda_3\mathcal{L}_1,
\end{equation}
where the term $\mathcal{L}_1$ is the Mean Absolute Error (MAE) \cite{10892224} employed to calculate the element-wise error. By following \cite{isola2017image, 10892224}, $\lambda_3$ is set to $100$.

\begin{figure}[t]
    \centering
    \includegraphics[width=0.95\linewidth, height=0.5\linewidth]{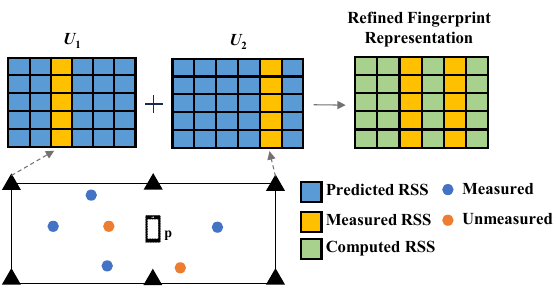}
    \vspace{-10pt}
    \caption{An example of the proposed localization procedure.}
    \vspace{-15pt}
    \label{fig:locpro}
\end{figure}

\vspace{-0.5cm}

\subsection{Predicted Fingerprint Optimization Procedure}
\label{sec:opt}

As discussed earlier, the proposed ComGAN framework predicts missing RSS values from incomplete samples and their associated RPs. For unmeasured RPs, a fingerprint subset is constructed by combining available measurements with the predicted RSS values. The overall procedure is illustrated in Fig.~\ref{fig:locpro}. In this work, one RU collects RSS measurements across multiple RPs, and the corresponding hybrid RSS samples form a subset denoted as $U$. The single-RU based configuration provides the flexibility to select the number of measured RUs according to available resources while maintaining prediction accuracy. To further enhance prediction robustness, $K$ RUs ($K < N$) are randomly selected, and each selected RU generates one subset. These subsets are then aggregated to form a refined fingerprint representation. Let $\mathbf{r}_{\text{meas}}$ be the measured entries, $\hat{\mathbf{r}}^{(k)}$ the predicted entries from the $k$-th subset, and $\mathbf{m}$ a binary mask indicating measured dimensions, the refined RSS measurement is given by:

\begin{equation}
\tilde{\mathbf{r}} = \mathbf{m}\odot\mathbf{r}_{\text{meas}} + (1-\mathbf{m})\odot \frac{1}{K}\sum_{k=1}^{K}\hat{\mathbf{r}}^{(k)}.
\end{equation}

\section{Experimental Results}
\label{sec:experimentalSetup}

This section evaluates the RSS completion scheme and demonstrates its improvements in fingerprinting localization.

\vspace{-0.5cm}

\subsection{Experimental Setups}
A practical RSS dataset is built based on real-world experiments conducted over long-term evolution (LTE) networks, where the details of the experimental setup are shown in \cite{11165772}. There is a total number of $40$ RPs distributed in the room, where each of them is within a $1$ meter distance, and $15,244$ RSS samples are collected. The RSS measurement data and the related RPs are split into three datasets, as shown in Table \ref{tab:data}. In addition, six RUs are evenly deployed within the localization area, determining the dimensionality of the RSS samples. To enhance the diversity of incomplete input data in the training dataset, five out of the six RU values in each RSS sample are randomly selected and replaced by zeros. 

The hyperparameters used in this work are empirically determined through preliminary experiments to balance model complexity and training efficiency. The generator comprises five Conv1D layers with 16, 64, 128, 64, and 16 filters, respectively, followed by a fully connected layer of 96 neurons for feature extraction. This configuration enables the network to learn both fine-grained variations and high-level structures in RSS data. Similarly, the critic contains three Conv1D layers with 16, 64, and 128 filters, followed by two fully connected layers with 8 and 2 neurons, which are sufficient to capture distributional discrepancies between real and predicted samples. To stabilize training, all input data are standardized to a normal distribution with zero mean and unit variance~\cite{10505767}, preventing scale differences among RUs from biasing the learning process. Both the generator and critic are optimized using Adam~\cite{kingma2014adam} with a learning rate of $1\times10^{-4}$. For the proposed optimization strategy, two fingerprint subsets from randomly selected RUs are employed. The localization network was a convolutional neural network (CNN) proposed by \cite{8662548}. All experiments are conducted on a CUDA-enabled NVIDIA Quadro RTX A5000 GPU with 16 GB of memory, ensuring efficient convergence within a reasonable time.

\begin{table}[t]
\centering
\caption{Dataset partition and statistics}
\setlength{\tabcolsep}{4.8pt}
\begin{tabular*}{\columnwidth}{|c|c|c|c|}
\hline
Dataset         & RSS Training & RSS Testing & Localization Testing \\ \hline
RP              & 15       & 15      & 10                   \\ \hline
RSS Measurement & 5731     & 5644    & 3869                 \\ \hline
\end{tabular*}
\label{tab:data}
\vspace{-15pt}
\end{table}

To evaluate RSS prediction performance, two complementary metrics are used: element-wise RSS error and localization error, both measured by the root mean squared error (RMSE). The element-wise error quantifies the accuracy of individual RSS reconstructions, while the localization error reflects the positioning accuracy, thereby indicating the practical impact of prediction quality. In addition, to enable comparisons across datasets with different signal ranges, the normalized RMSE (NRMSE) \cite{9684959} is used to provide a scale-invariant assessment of reconstruction accuracy. 

\vspace{-0.3cm}

\subsection{The Performance of the RSS Completion and Localization}
\label{subsec:results}
An element-wise RSS error analysis is first conducted to evaluate ComGAN’s reconstruction accuracy by comparing predicted RSS data with ground-truth measurements. This analysis examines the precision of individual RSS components rather than aggregate statistics. Across the testing dataset, the average RSS error is 7.67 dBm for RMSE and 0.15 for NRMSE. To assess statistical reliability, a non-parametric bootstrap resampling method~\cite{tibshirani1993introduction} is applied, where RMSE values are resampled 1,000 times to obtain a 95\% confidence interval. The resulting interval of [7.61, 7.85] dBm indicates that the reported performance is statistically stable.

To further evaluate the quality of the predicted data, the localization performance of the CNN-based localizer is compared across three training datasets constructed from the ComGAN testing set. Each dataset contains the same RPs but differs in RSS completeness: (i) incomplete RSS samples, (ii) fully measured RSS data, and (iii) hybrid samples in which unmeasured RSS values are predicted by ComGAN and refined using the optimization method. When trained on the first dataset, the CNN yields a localization error of 4.32 m, compared with 0.89 m using the complete measured data. The hybrid dataset achieves a similar error of 0.93 m, indicating that ComGAN-predicted RSS values can effectively substitute real measurements without performance degradation. These results suggest that ComGAN can substantially reduce data collection requirements while maintaining localization accuracy comparable to that of fully measured data. Furthermore, on a 12th Gen Intel\textsuperscript{R} Core\textsuperscript{TM} i7-1265U CPU, the average inference time per hybrid sample generation is approximately 0.011 s. These results indicate that ComGAN can perform low-latency inference even on resource-constrained hardware, supporting its applicability in real-time or near-real-time IoT localization systems.

\begin{table}[t]
\centering
\caption{Comparison of localization error calculated based on the ablation study.}
\setlength{\tabcolsep}{12.1pt}
\begin{tabular*}{\columnwidth}{|c|c|c|c|c|}
\hline
\textbf{Model}     &\multicolumn{2}{|c|}{\textbf{ComGAN}}  &\multicolumn{2}{|c|}{\textbf{Lite-ComGAN}} \\ \hline
\textbf{\begin{tabular}[c]{@{}c@{}}Optimization\\ Method\end{tabular}} & With   & Without & With        & Without \\ \hline
\textbf{\begin{tabular}[c]{@{}c@{}}Localization\\ Error (m)\end{tabular} } &   0.93     &  1.03       &    0.96         &  1.08       \\ \hline
\end{tabular*}
\label{tab:ab}
\end{table}

\vspace{-0.5cm}
\subsection{Ablation Study}

To evaluate the contribution of the proposed components in ComGAN, three ablation studies are conducted. The first study examines the role of the auxiliary sub-network, which enforces spatial consistency between RPs and RSS values to guide feature learning. A simplified variant, Lite-ComGAN, is constructed by removing this sub-network and its regularization term $\text{MSE}(\hat{c}_1,\hat{c}_2)$, while keeping all other settings unchanged. On the ComGAN testing datasets, Lite-ComGAN yields an average element-wise error of 7.77 dBm (NRMSE 0.16), slightly higher than 7.67 dBm with the full model. In addition, the proposed optimization strategy is also examined. Table~\ref{tab:ab} shows that both ComGAN and Lite-ComGAN achieve lower localization error values when the optimization is used, confirming the effectiveness of the optimization method. Although the difference in RSS prediction is minor, Lite-ComGAN consistently leads to higher localization errors, confirming that the auxiliary sub-network provides useful structural guidance for RSS reconstruction. The second study validates the effect of the number of subsets $K$. Without optimization ($K=1$), the localization error in terms of RMSE is 1.03 m. With $K=2$, the RMSE drops to 0.93 m, and with $K=3$, it further decreases slightly to 0.92 m. Since the improvement from $K=2$ to $K=3$ is marginal, we adopt $K=2$ as a practical trade-off.

\begin{figure}[t]
    \centering
    \includegraphics[width=0.9\linewidth, height = 0.77\linewidth]{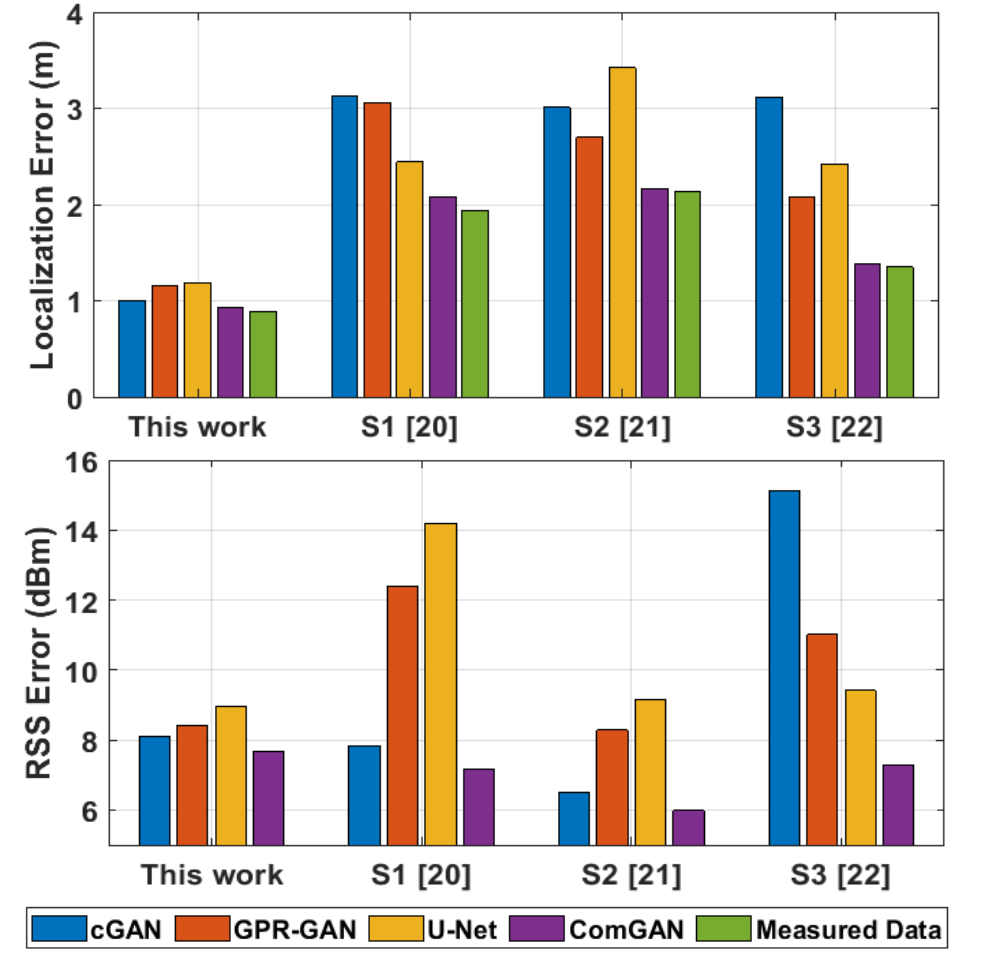}
    \caption{Performance comparison of localization error calculated based on the ablation study.}
    \label{fig:compbench}
    \vspace{-20pt}
\end{figure}

\vspace{-0.5cm}
\subsection{Benchmarking and Adaptability Analysis}
To assess the performance of the proposed scheme, several state-of-the-art methods are used for comparison. For GPR-GAN \cite{zou2020robot}, the GPR part keeps the same, but the random vector in the following GAN used in \cite{zou2020robot} is replaced by incomplete RSS samples, forming the input composed of both incomplete and coarse RSS observations. Considering the incomplete RSS samples as noisy measurements, a U-Net proposed in \cite{11016042} is also compared. In addition, given that ComGAN is inspired by the cGAN framework, the cGAN model \cite{boulis2021data} is also included in the comparative analysis. The cGAN model takes the same input and output data as ComGAN. The training and testing configurations follow the setup described previously. Performances are also evaluated using localization error values and element-wise error values.

To investigate the adaptability of the ComGAN concurrently, three additional scenarios \cite{11016042, feng2022analysis,yuen2022wi} are also evaluated, where samples from them are also split into RSS training, RSS testing, and localization testing sets following the same RP-based ratio as shown in Table \ref{tab:data}. This three scenarios are labelled as: S1 \cite{11016042}, S2 \cite{feng2022analysis}, and S3 \cite{yuen2022wi}. By following the description in Section~\ref{subsec:results}, for each scenario, three distinct training datasets are constructed to evaluate the localization performance. The results, including the scenario of this work, are shown in Fig. \ref{fig:compbench}. As shown in Fig. \ref{fig:compbench}, ComGAN delivers the lowest element-wise prediction error. It also supports the CNN localizer in achieving a localization error that most closely matches that of real RSS data. These results indicate that ComGAN is capable of accurately predicting RSS data in different environments, showing the adaptability potential to more complex scenarios.

\section{Conclusion}\label{sec:conclusion}

This paper presented a ComGAN, an adversarial framework for predicting missing RSS measurements from incomplete samples and their associated RPs. By integrating adversarial training, hybrid losses, a spatial-dependency sub-network, and a reconstruction refinement strategy, ComGAN consistently outperforms state-of-the-art methods, achieving the lowest RSS reconstruction and localization errors across diverse real-world scenarios. These results confirm its effectiveness, adaptability, and practical value for fingerprinting localization.

\newpage
\bibliographystyle{IEEEtran}
\bibliography{bib}

\end{document}